# Physical and chemical modifications of polymeric surface for enhanced epithelial cells adhesion

Laura M. S. dos Santos[1], Jonathas M. de Oliveira[2], Sendy M. S. do Nascimento[1], Artur F. Sonsin[1, 4], Vitor M. L. Fonseca[3], Juliane P. Silva[3], Emiliano Barreto[3*], Cléber R. Mendonça[5], Alcenísio J. Jesus-Silva[1], Eduardo J. S. Fonseca[1*]

[1] Optics and Nanoscopy Group, Institute of Physics, Federal University of Alagoas (UFAL), 57072-970, Maceió, Alagoas, Brazil
[2] Federal Institute of Alagoas (IFAL), 57230-000, Coruripe, Alagoas, Brazil
[3] Laboratory of Cell Biology, Institute of Biological Sciences and Health, Federal University of Alagoas (ICBS/UFAL), 57072-970, Maceió, Alagoas, Brazil
[4] Federal University of Rondônia (UNIR), 76940-000, Rolim de Moura, Rondônia, Brazil
[5] Institute of Physics of São Carlos, University of São Paulo, CP 369, 13560-970, São Carlos, São Paulo, Brazil

*Corresponding author: eduardo@fis.ufal.br; emilianobarreto@icbs.ufal.br

**Abstract**

In tissue engineering, 3D scaffolds and chemical treatments are often used for providing a cell-friendly surface for improving cell adhesion and tissue growth. Indeed, the cell adhesion degree can be controlled by physical-chemical changes in the surface of substrates, such as wettability, surface charges and roughness. In this work, we describe the synthesis, characterization and cytocompatibility of photoresins useful for construction of cell scaffolds via two-photon polymerization. Additionally, we have demonstrated a simple surface treatment method that promotes cell adhesion. This method alters the surface charge of the polymer and enhances the adhesion of epithelial cells. Our results indicate an efficient approach for modifying the surface of biocompatible polymer scaffolds with the purpose of enhances the performance of cell functions suitable for tissue engineering and regenerative medicine.

## 1. Introduction

Polymeric materials are at the forefront of the development of biophotonic and biomedical devices. These materials have numerous applications, including biosensors for diagnosis, implants, and 3D environments suitable for cell culture [1-3]. The attractive features of polymers include ease of synthesis, large scale production, capacity for functionalization with nanoparticles and others chemical functional groups, as well as their interesting optical and mechanical properties [4, 5]. Furthermore, they can be synthesized to present adjustable flexibility, which makes them an excellent tool for drug delivery and as scaffolds for cells and tissue growth [6-9]. In particular, the acrylic-based polymers are widely used in biomedical application due to good biocompatibility, biodegradability, and mechanical properties [10-13].

Cell adhesion to a surface is crucial for various cellular functions, including proliferation and differentiation. It also plays a significant role in the development of biomaterials and implantable sensors or devices [14]. The understanding of the factors that affect the cell-surface adhesion allow to control of the cell behavior in the local environment surrounding to benefit the investigation on the tissue engineering technologies [15]. Indeed, the relationship between cellular behavior and the wettability of polymer surfaces has been investigated in literature, in particular those related to cell adherence [16]. Although the studies have suggested an improvement of cell adhesion on hydrophilic surfaces there are other factors that must be considered for a complete understanding of the cell behavior such as surface energy, roughness and porosity of polymer surfaces [17-20]. For instance, B. Majhy and colleagues [21] showed that cervical and breast cancer cell lines present different behavior for the same substrate, adjusting their adherent properties for different roughness ratio. This demonstrated that each cell lineage response in a particular way to a specific feature of the substrate. Indeed, in cell-material interactions, both physical and chemical features need to be adjusted to create a comfortable environment for the cells [22].

Numerous approaches for surface modification have been used to alter the materials properties and enhances the cell attachment, among them include plasma treatment, chemical vapor deposition, dynamic surface modification, protein adsorption, and salinization [23, 24]. Despite of great diversity of methods able to induce modification of polymer surface [25-27], there is still a need to propose cheaper and faster methods for treatment of surfaces. Furthermore, it is important to ensure that the biocompatibility of each new polymeric material is paired with a suitable surface treatment. This approach helps create a surface that is more favorable and biologically similar to the organism. Following an eco-friendly trend, the use of phosphate buffer saline (PBS) solution to induce surface modifications in inorganic materials meet the currently requirements for facilitating a reduction of the organic reagents' usage and less waste production [28, 29]. PBS is commonly used to simulate biological solutions in research and laboratory settings. It is chosen for its ability to closely match the osmolarity and ion concentrations found in the human body. As an isotonic and non-toxic medium, PBS provides a suitable environment for maintaining the viability and functionality of living cells. While the technique of increasing the surface hydrophilicity of polymers using PBS is well-documented [30-36], the effects of PBS on the roughness and appearance of functional groups on the surfaces of polymers based on pentaerythritol triacrylate (here referred to as P59) and trimethylolpropane ethoxylate triacrylate (here referred to as S59) in creating a cell-friendly surface have not been extensively explored and described, to the best of our knowledge.

In this work, we conducted an evaluation of the adhesion of human epithelial cells A549 on acrylated-based photoresins. We further investigated the impact of chemical modification induced by PBS on cell adhesion on polymeric surface. Our results were supported by several characterization methods.

## 2. Material and methods
### 2.1 Materials

The monomers used for preparation of photoresins were pentaerythritol triacrylate and trimethylolpropane ethoxylate triacrylate. The photoinitiator 2-Hydroxy-4′-(2-hydroxyethoxy)-2-methylpropiophenone (Irgacure® 2959) was used to initiate photochemical crosslinking. All chemicals were purchased from Sigma-Aldrich.

**2.2 Preparation and characterization of the polymeric substrates**

Two polymeric samples were prepared using a combination of each monomer and 1% by mass of photoinitiator, designated P59 (pentaerythritol triacrylate) and S59 (trimethylolpropane ethoxylate triacrylate). The photoresins were gently mixed for 24h at room temperature.

The substrates were molded in circular blades (16 mm diameter × 0.6 mm thickness) and polymerized by UV light for 20 minutes. Afterwards, the samples were carefully immersed in ethanol for 20 minutes to wash away the unpolymerized resin. This process was repeated 3 times to ensure complete removal.

A strategy to make substrates more hydrophylics and potentially more adherents for cells was to treat the polymeric surfaces with phosphate-buffered saline (PBS 1×, pH 7.4). A PBS solution consisting of NaCl, KH2PO4 and Na2HPO4 was prepared. After that, the polymeric substrate treated for 60 minutes at room temperature. Here, the control is commercial adherent cell culture plates treated for promoting cell adhesion. Finally, the samples were dried in an oven at 50º C for 10 minutes. After the PBS treatment, the samples were renamed P59/PBS and S59/PBS.

**2.3 Contact angle measurements**

The investigations of wetting phenomena on the substrate surface were made using the sessile drop method, an accurate and reproducible optical method for contact angle measurements. Only one drop of water (3μL) was deposited on different samples. The contact angle measurement was performed using Theta Optical Tensiometer (T200 Biolin Scientific) with a embeded CCD camera and Attension software. For wetting phenomena, all 5 measurements for each sample were carried out at 23 °C.

The interactions between solid and liquid are important in various processes as they determine the adhesion between the phases. Solid-liquid interactions are determined by the surface free energy (SFE) of the solid and the surface tension of the liquid applied. The SFE method used in this work was Owes, Wendt, Rabel and Kaelble model. Uses geometric mean to treat the molecular interactions. The idea of it is dividing the SFE into individual components. We use only polar interaction with films. The polar liquid chosen was water due its large polar component, availability and non-toxic nature. The SFE polar ($\gamma_p$) was calculated using equation which is given as:

$$\gamma_p = 0.5 \cdot \gamma_{lv}(1 + cos\theta), \qquad (1)$$

where $\gamma_{lv}$ is the superficial tension of polar liquid, in this case, water (72.8 $m$Jm$^{-2}$) and $\theta$ is the angle forming in contact angle measurement.

**2.4 Atomic force microscopy (AFM)**

All measurements were obtained using a standard AFM setup (Multiview 4000 ™, Nanonics, Israel), with a combined optical microscope (BXFM, Olympus, Japan). The AFM system was acoustically isolated to reduce any interference by ambient noise during the measurements, and the instrument was secured on an active damping table to suppress mechanical noise. The topography of the substrates was imaged (256×256 pixels) in tapping mode with a scan rate of 0.3-1.0 Hz in an area of 15×15 µm². About nine regions were analyzed for each substrate. The roughness of the samples Control (CTRL), S59, P59 and treated with PBS was analyzed using software WSxM, through the mean parameters of the roughness of each sample. For each image, two areas of 7.5×7.5 µm² were selected and the average roughness ($R_a$) was calculated as the absolute mean of the heights of the irregularities along the profile.

$$R_a = \frac{1}{N}\sum_{i=1}^{N}|z_i - \bar{z}|, \qquad (2)$$

where $N$ is the number of sample points, $z_i$ is the height of each sample point and $\bar{z}$ is the average height of the sample points.

Data expressed as mean ± standard deviation (SD). To show that all data were normally distributed, the Kolmogorov-Smirnov and Shapiro-Wilk tests was performed at the 0.05 level, with a population normal.

**2.5 Fourier transformation infrared (FTIR) and UV–Vis spectroscopies**

The infrared spectra (FTIR) of samples were obtained, at room temperature, using an IRPrestige-21 spectrophotometer (Shimadzu, Kyoto, Japan) coupled with an attenuated total reflectance (ATR) accessory with ZnSe crystal. The spectral range was 4000–800 $cm^{-1}$, 4 $cm^{-1}$ resolution and 120 scans. The band intensities were expressed in transmittance (%) with a diffuse reflectance accessory (DRS-8000). The FTIR spectra of each sample were obtained before and after the PBS treatment. The absorption spectra of the samples were performed by an UV-3600 spectrophotometer (Shimadzu, Kyoto, Japan). The spectra were recorded in the range of 300–700 nm.

**2.6 Cell Culture**

Human alveolar epithelial cell line A549 were used in this study and grown in Dulbecco Modified Eagle's Medium (DMEM) supplemented with 10% fetal bovine serum (FBS), 1.467% of L-glutamine, 1.4% of glycose, 1% sodium pyruvate and 0.02% of penicillin/streptomycin and maintained at 37°C in a 5% $CO_2$ humidified atmosphere. Cells were passaged every 2 days with medium changed.

Prior to cell adhesion studies with the polymers of interest, the well-dishes culture was covered with 1 mL of polymeric substrates or PBS. After 1h, the overage liquid was withdrawn and residual humidity was dried in a heating incubator for 5 minutes. Next, the 12-well plate was transferred to a laminar flow hood for 1 hour under UV light.

**2.7 Cell seeding and adhesion**

Cells were detached from their culture dishes by incubating with a solution of 2.5 g/L trypsin and 0.38 g/L of ethylenediaminetetraacetic acid (EDTA) for 5 min. The dissociated cells were seeded ($3\times10^4$ cell/well) on the polymer surface and maintained in supplemented DMEM medium under culture conditions for 24 hours. Photographic register was taken at 0 h and 24 h to evaluate the morphological differences related to the adhesion.

**2.8 Quantification of cell adhesion**

The interaction of epithelial cells after contact with polymeric substrates, before and after PBS treatment, was observed by optical microscopy (100× magnification). The cells images were captured in two different regions of the substrate. In each picture, we selected 4 regions with the same area (~ 57×57 µm$^2$). For each specific region, the cells were counted and its morphology was analyzed by Gwyddion software (version 2.3) and the resulting data were processed with Origin (OriginLab). The cell adhesion quantification was guided by the morphological aspect of cells evoked by adhesive behavior after interaction with polymeric substrates. We have considered that non-adherent cells present a rounded and shiny appearance, while adhered cells appear more elongated with spindle shaped, fibroblast-like morphology (Figure 1).

Quantitative analysis of the morphology was carried out for better understanding of the changes in adhesion-induced cell shape. Parameters such as Shape Factor and cellular aspect ratio of the cells were calculated. The Shape Factor $\phi$ (ranging from 0 to 1) relates the ratio between the cell surface area and the cell perimeter [37, 38]. The cellular aspect ratio is approximately the ratio of the minor axis (width) to the major axis (length) of an ellipse fitted in cell [39]. Therefore, through these parameters it was possible to indicate how circular the cells are. This means that, for more rounded cells, the area-perimeter ratio increases, as well as the relationship between cell axes, making the Shape Factor and cellular aspect ratio closer to 1. While, for cells that are adhered to the substrate and acquire elongated geometries, these parameters are less than 1. Cell morphological parameters were calculated for at least 60 manually labeled cells for each experimental condition. Experimental data were expressed as mean ± standard deviation.

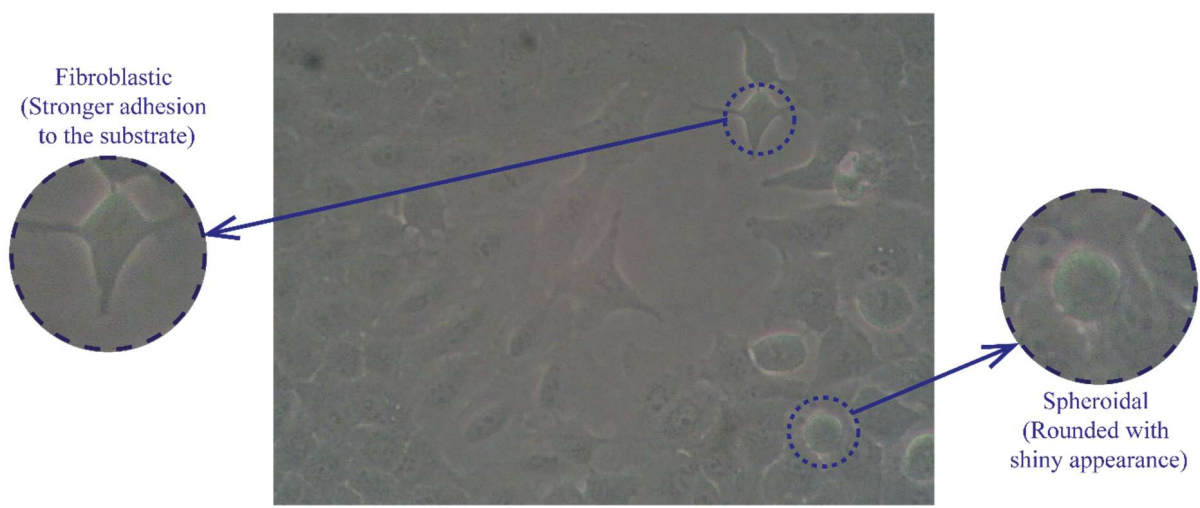

**Figure 1: Light microscope images of A549 cells grown in DMEM medium for 24 hours commercial adherent cell culture plates (100× magnification). Cells identified by shows spheroidal or fibroblast-like morphology were enlarged in the insert pictures.**

## 3. Results

As shown in Figure 2, the polymers P59 (pentaerythritol triacrylate) and S59 (trimethylolpropane ethoxylate triacrylate) were analyzed using UV-vis spectroscopy. The characteristic features of UV–Vis absorbance spectra revealed that the onset of absorption was identical before polymerization. The Irgacure® 2959 was responsible for the predominant absorption band below 350 nm and a transparency for visible radiation in both samples (Figure 2).

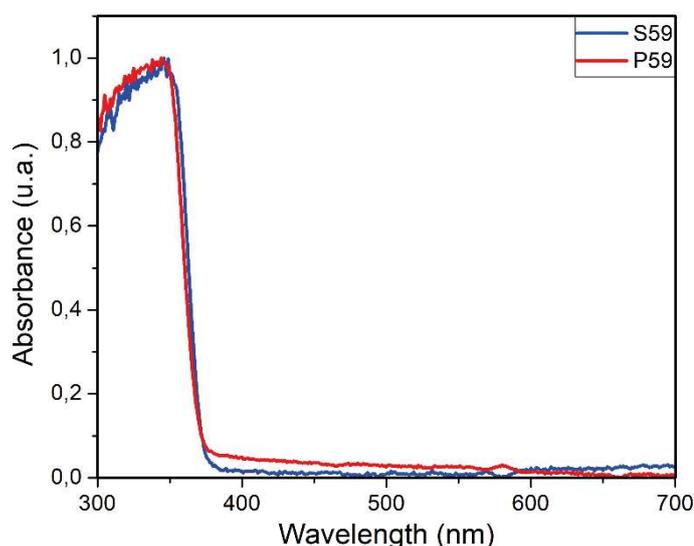

**Figure 2: UV–visible absorption spectrum the polymeric samples P59 and S59.**

From this point forward, P59 and S59 were utilized as substrates to promote cell adhesion. For this purpose, the cells in fibroblastic (elongated) or rounded shape after 24 h of exposure to polymeric surfaces were quantified. In addition, in another set of experiments, we used PBS to induce changes on surface of polymers P59 and S59 before allow cells interaction. The Figure 3A shows the percentage of adhered cells after interaction with each sample treated or not with PBS.

By observing the adhesion of cells, and comparing with the control plates, it was possible to note that S59 and P59 did not presented enough physicochemical features to promote the cell adhesion. After treatment with PBS on surface of the samples, the S59/PBS substrate allowed an increased in the adhesive behavior of cells about 3 times than compared to condition without treatment with PBS (Figure 3A). The cells behavior on S59 substrates after PBS-treatment were comparable to behavior presented on commercial control plates. The P59 polymer did not induced changes in cell adhesive behavior after PBS-treatment (Figure 3A). These results, revealed that S59 substrates after PBS-treatment showed a surface cell-friendly to adhesion of epithelial cells.

Figure 3B shows the results for water wettability measurements, conducted before and after the PBS treatment. The contact angle of all samples exhibited a reduction of approximately 20%, indicating an increase in hydrophilicity due to the PBS treatment (Figure 3B). Notably, the P59 samples demonstrated a greater decrease in contact angle, shifting from 64.93° to 41.48° following the PBS treatment (Figure 3B).

The surface free energy was calculated via equation (1) and related to the cell adhesion in Figure 3C. We can observe that there was a correspondence between the free energy and the cell binding to substrate. It is important to note that the control (CTRL) substrates used in this study are commercial polymeric plates treated for promoting cell adhesion. For surface free energy (SFE) values close to the CTRL substrate, E = 42.78 $mJ/m^2$, the percentage of adhesion increases. Therefore, the cell adhesion is bigger on S59/PBS substrate compared to P59/PBS substrate.

The polar component of SFE showed a correspondence with cell adhesion. The CTRL substrate had the lowest polar SFE value, indicating the best cell adherence. The S59 PBS-treated substrate had a higher polar SFE of approximately 57 $mJ/m^2$, which also promoted cell adhesion. However, a larger SFE (64 $mJ/m^2$), as observed in the P59 PBS-treated substrate, resulted in lower cell adhesion. These results suggest that the polar SFE plays a significant role in cellular adhesion.

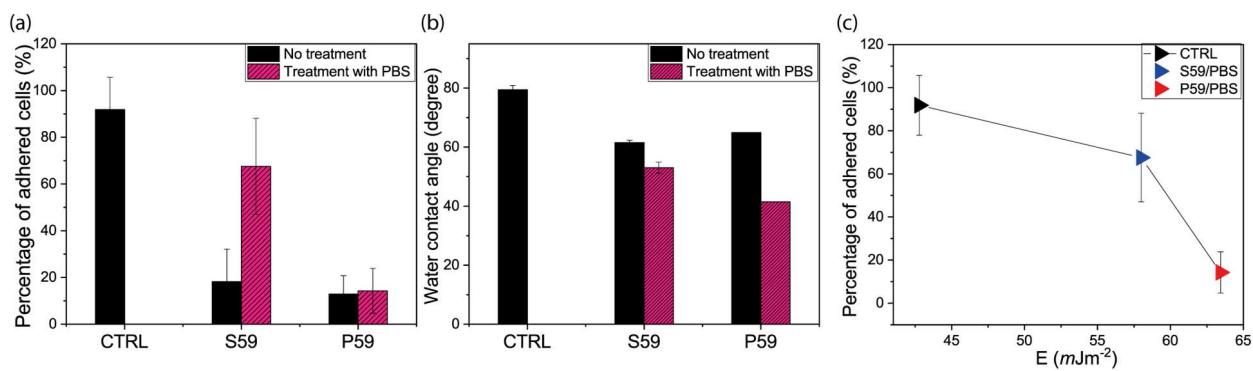

**Figure 3: Evaluation of the adhesive behavior of cells (A), wettability measurements (B) and free energy (C) of polymeric surfaces treated or not with PBS.**

As topography of surface is an important environment component able to influence the cell adhesion, we focused in evaluate the roughening of the polymeric surfaces treated or not with phosphate buffer (PBS) to regulate cell adhesion. To achieve this objective, the AFM system was utilized to observe the topographies of the polymeric substrates, allowing for the investigation of surface nanoroughness. Figure 4A shows the average roughness ($R_a$) of the polymeric substrates before and after the PBS treatment.

As shown in Figure 4A, the average roughness of the all samples was modified by the PBS treatment, making P59 samples less rugged but increasing the roughness of the S59 samples. Interesting enough, the PBS buffer adjusted the average roughness of all surfaces for about 5.30 nm. The normality test showed the population means are significantly different, with 0.05 level.

We analyzed a transversal cutting of the polymeric plans (Fig. 4B and 4C). In these Figures, we can observe that S59/PBS substrate is notably less wrinkled than P59/PBS samples. Thus, the roughness surface ratio r, defined as the ratio between the actual surface area and the projected solid surface area, can give us more accurate clues about the behavior of the cells on each substrate.

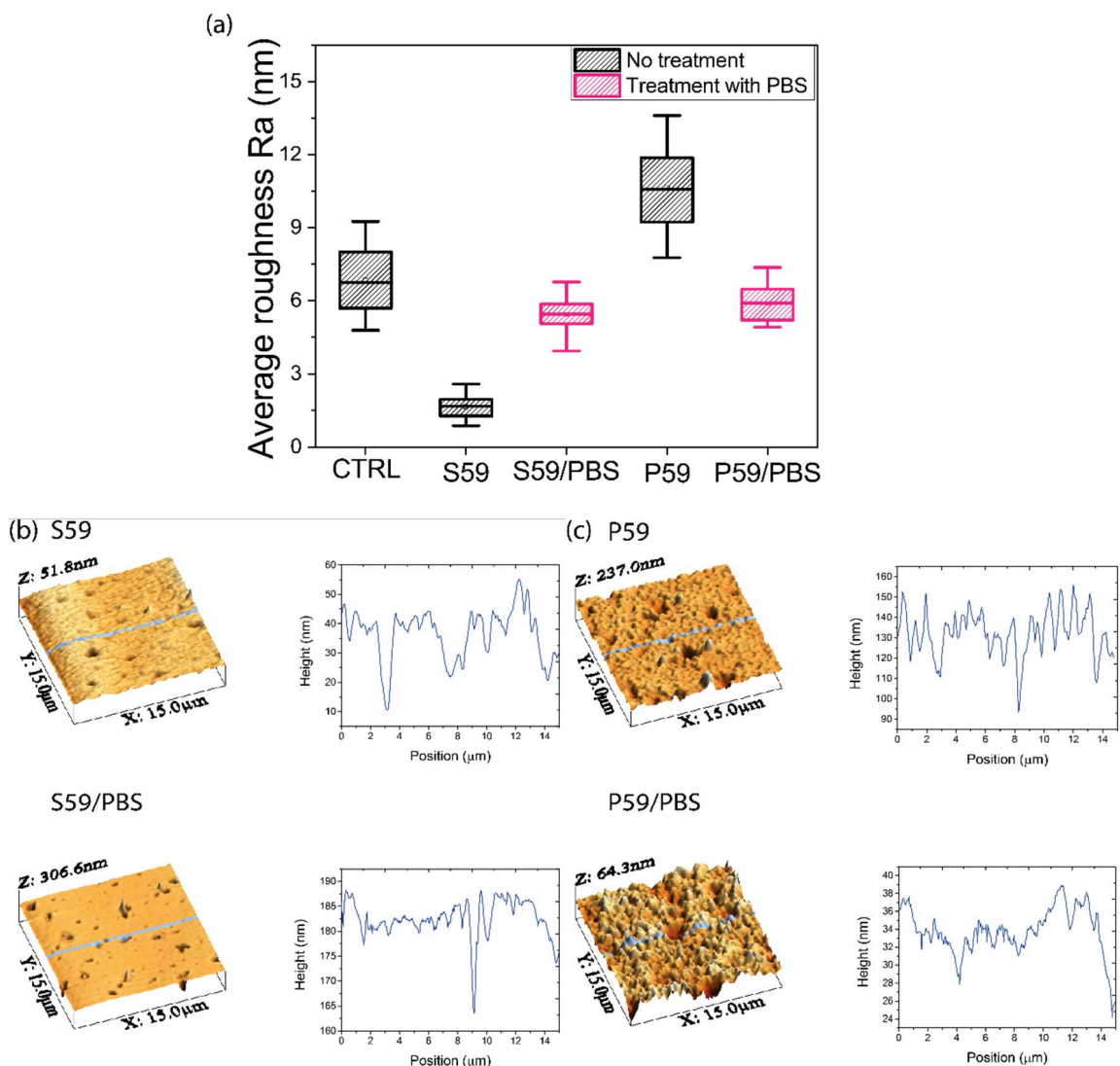

Figure 4: (a) The roughness average ($R_a$) values of the samples CTRL, S59 e P59 and treated with PBS. AFM topographic images detected through atomic force microscopy of the sample substrates, namely (b) S59, S59/PBS, (c) P59 and P59/PBS.

Indeed, Figure 5 relates the roughness ratio (r) and the percentage of adhered cells on each substrate. We can observe that CTRL samples present an interesting roughness profile for the cell behavior, with r = 1.0075. Such topography pattern was also observed by S59/PBS samples, with r = 1.0052, in such a way that the counting of adhered cells resembled to seeded cells on CTRL substrates. However, in P59/PBS samples r = 1.0138, making the topography wrinkled enough for discouraging the cell's adhesion properties. It is important to keep in mind that the roughness ratio is just a reference number, representing a convergence point between a specific cell and a particular nano-topography. Even so, the roughness ratio (r) should be considered as a more significant parameter than average roughness ($R_a$), insofar as quantifying the nano-wrinkling of the surface. Indeed, the cellular behavior face the nano-roughness of the substrate can be expressed by its capacity of fixation on the substrate and subsequent morphologic change.

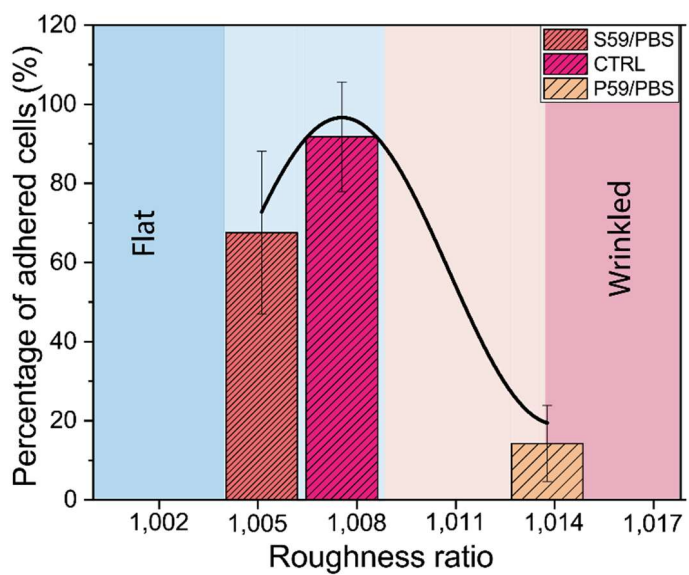

**Figure 5:** Effect of roughness ratio of polymeric films treated with PBS on the fixation behavior of the A549 cell.

Next, Infrared spectroscopy was employed to localize the chemical changes occurring in the functional groups of the polymeric substrates treated with PBS. Figure 6 shows the recorded FTIR spectra of untreated polymeric samples (black lines), PBS-treated samples (blue lines), PBS buffer sample (dark yellow lines) and the difference spectrum (pink lines) between the surfaces before and after PBS treatment. It is very clear that the S59/PBS samples were affected by the phosphate buffer treatment, unlike P59/PBS samples. Observing the pink lines (Figure 6A and 6B) we can see some important vibrational bands excited by the PBS treatment on S59 substrate but not observed on P59 substrate.

In spectra were observed phosphates bands, attributed to PBS treatment. However, other bands were shown for S59 substrate, indicating interactions with PBS. Mainly, formation of oxygenated and hydroxylated groups, as OH bends at 1418 cm$^{-1}$, the peaks at 1760 cm$^{-1}$ and 1714 cm$^{-1}$ depicted the formation of functional groups alkyl carbonate and carboxylic acid, respectively. The peak 1232 cm$^{-1}$ owed to the stretching vibrations of aryl-O. The peak 2972 cm$^{-1}$ is related to methylene. Draws attention the vibrational modes centered in 1370 cm$^{-1}$ (hydroxyl bond (–OH)) and 1086 cm$^{-1}$ (stretch P-O dos phosphate ion), this last one is specifically related to H$_2PO_4^-$ or HP$O_4^{2-}$ bond, suggesting the formation of a thin phosphate layer on the S59 surface after adsorption process.

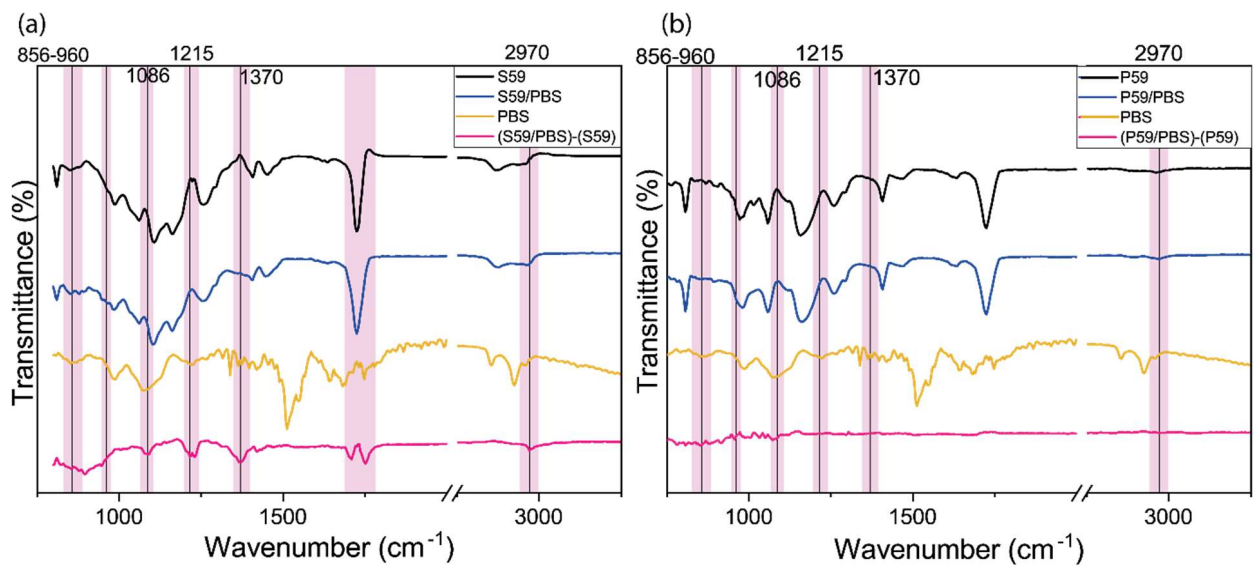

**Figure 6: FTIR characterization spectra of the samples and difference spectrum of the films polymerics (film original minus film treatment with PBS).**

Figure 7A shows the values of the measured morphological parameters for a better analysis of cell adhesion to surfaces treated with PBS. For the CTRL samples, the cells exhibited a shape elongated with a larger cell area, which resulted in a Shape Factor and aspect ratio value of 0.56 ± 0.06 and 0.32 ± 0.07, respectively, which indicates high cell adhesion. Our analysis showed that the S59 substrate treated with PBS increased the adhesion of A549 cells. The Shape Factor and aspect ratio values for these samples were 0.58 ± 0.05 and 0.34 ± 0.06, the closest to the control, respectively. For the P59/PBS samples, the values of these parameters were closer to 1. This indicates the lower cell adhesion to the surface. The cell behavior quantified and revealed in Figure 3A can be observed in Figure 7B by the analysis of the cell morphology. We can see clearly how the S59 substrate promotes both the cell adhesion, as the CTRL plates, and the cell proliferation, as the P59 substrates. Indeed, two required cell behavior when developing biomaterials for tissue engineering.

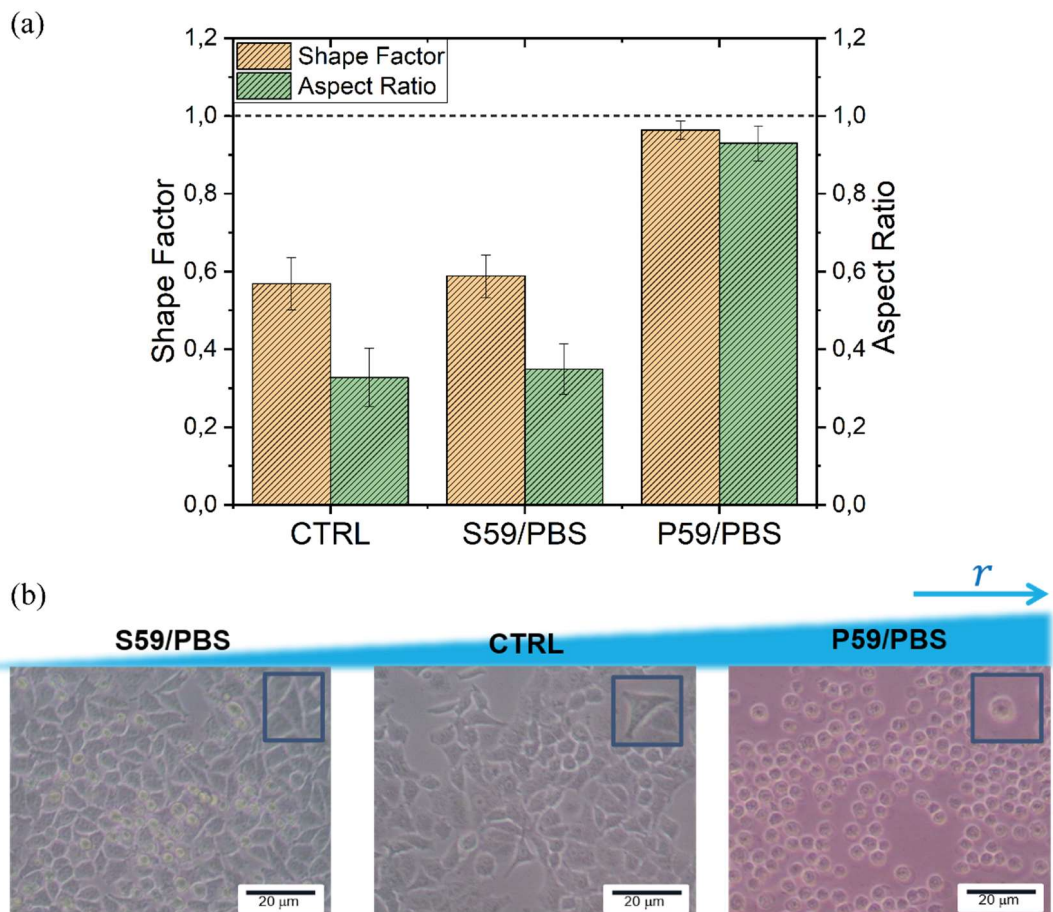

**Figure 7: (a) The value of the shape factor and aspect ratio of A549 cells after the surface modification with PBS and CTRL. (b) The experimental images of A549 cells cultured on different surfaces for 24 h.**

## 4. Discussion

The cell-substrate interaction is strongly influenced by physicochemical properties, including wettability, roughness, and surface free energy. Consequently, to control these characteristics allows predicting and improving biological activities, including cell adhesion. Polymers have been widely used as a substrate for 2D cell cultivation and in the fabrication of 3D cell scaffolds [40-42]. Although acrylic-based polymers have numerous advantages, they are materials with physicochemical characteristics that can inhibit cell adhesion [43, 44]. Since each cell type has specific adhesion properties, it is necessary to adjust the parameters present in the substrate to the cell behavior, signaling adequate conditions for spreading.

Previous studies have already investigated the modification of the polymeric surface roughness by the PBS buffer together with thermal treatment [45]. Unlike previous studies, here we performed the treatment with PBS on polymeric surfaces S59 and P59 was demonstrated without the need for further heat treatment, and it had a dominant effect on the adhesion of pulmonary epithelial cells. This occurred because the physicochemical properties related to wettability, surface energy, and roughness ratio are modified after the interaction with PBS, being characteristics already known to influence cell morphology and adhesion [23, 46, 47].

In our *in vitro* studies to determine cell adhesion, we demonstrated that the cells showed higher adhesion and elongation after 24 hours on the S59/PBS substrate. Under this condition, the values of shape factor and the cellular aspect ratio were closer to the control (S59/PBS: 0.56 and 0.32; CTRL: 0.58 and 0.34, respectively), reflecting the morphological confirmation of the cells. However, for the P59/PBS samples, the shape factor and the cellular aspect ratio values approached to 1 (0.96 and 0.92, respectively), resulting in less cell adhesion to the material's surface.

From there, we verified how these morphological changes were related to the roughness ratio on the polymer surfaces. Indeed, this is a crucial parameter to determine the effectiveness of cell adhesion, as it describes the actual interaction between the cell and the rough surface [48]. Once again, S59/PBS and CTRL coincided and presented a moderate roughness ratio ($1.005 \leq r \leq 1.008$), allowing the cell cytoplasmic membrane to conform to the grooves of the rough surface, promoting maximum interaction and stability between cell and substrate, resulting in better adhesion. On the other hand, P59/PBS presented a higher roughness ratio ($r \geq 1.014$), making it difficult for the cell cytoplasmic membrane to adapt to irregularities and fully reach all surface grooves. As a result, the cells were located on top of the rough substrate's protrusions without touching the bottom of the grooves. This reduced contact interaction between cell and substrate has lead to a significant decrease in cell adhesion, phenomena that finds support in previous works [24].

Furthermore, through FTIR analysis, we demonstrated that the PBS treatment was effective in functionalizing the polymeric substrate S59. The FTIR spectrum of these samples showed the presence of hydroxyl (OH) and carboxylic acid (COOH) functional groups, essential for promoting cell adhesion. Several studies focused on the biological activity of polymers indicate that the carboxyl and hydroxyl groups, along with the surface roughness promoted in the substrate, have the potential to promote excellent cellular affinity [21, 49-51].

Also, it is already known that adjusting the surface free energy can influence cell adhesion [52, 53]. Taking as a reference a commercially treated polymeric substrate to improve cell adhesion with a surface free energy of 42.78 mJ/m², we observed that the S59/PBS substrates approached their surface free energy to that of the control material (57.0 mJ/m²), which did not occur with the P59/PBS substrate (63.44 mJ/m²). This result is strongly linked to the fact that surface energy controls the material's wettability, which, in turn, can affect the number of proteins adsorbed during the interaction of cells and the polymeric substrate [54, 55].

These results indicate that we can design and develop 3D cell scaffolds and further adjust the environmental conditions through PBS treatment to promote cell adhesion. This optimization enhances the microfabrication process and facilitates the study of cellular behavior, as we can follow the basic protocols of biological assays, namely: scaffold fabrication through two-photon polymerization, removal of non-polymerized resin, treatment with PBS, sterilization, and cell culture.

## 5. Conclusions

The effect of PBS treatment as a method for modifying the surface properties of polymeric films made from S59 presents unique advantages studied here, such as moderate roughness, increased hydrophilicity, and the emergence of functional groups. This makes it an excellent candidate for controlling cell functions and tissue regeneration. The findings of this study open up new possibilities for including PBS treatment as a surface adaptation method to regulate cell responses, among the various surface treatment techniques already known, due to being an extremely simple, fast, eco-friendly, low-cost, and efficient process with a green footprint. The tendency of PBS treatment to improve the adhesion of A549 pulmonary epithelial cells to the S59 polymer is of significant relevance as it allows for the applicability of polymeric materials that were previously unsuitable for cell cultivation, as they did not enable the cells to exhibit their cellular behavior. Our study and conclusions were based on optical observations, spectroscopic characterizations, and mechanical analyses involving the cells and the substrate surface. Our results demonstrate that it is possible to regulate the physicochemical properties of the polymeric substrate with PBS treatment to enhance cellular responses.

**Funding:** The present research was supported by Fundação de Amparo à Pesquisa do estado de Alagoas – FAPEAL (APQ2019041000017), São Paulo Research Foundation – FAPESP (2019/25164-2, 2018/11283-7), Conselho Nacional de Desenvolvimento Científico e Tecnológico – CNPq, Coordenação de Aperfeiçoamento de pessoal de Nível Superior – CAPES (Finance Code No. 001).

**Conflict of Interest:** The authors declare that they have no known competing financial interests or personal relationships that could have appeared to influence the work reported in this paper.

Supervision, Validation, Writing–review, Funding acquisition; **Cléber R. Mendonça:** Validation, Visualization, Writing–review, Funding acquisition; **Alcenísio J. Jesus-Silva**: Validation, Visualization, Writing–review; **Eduardo J. S. Fonseca:** Conceptualization, Supervision, Project administration, Writing–review, Funding acquisition.

**Data Availability:** The data that support the findings of this study are available upon the request from the authors.